\documentclass[conference]{IEEEtran}

\usepackage[pdftex]{graphicx}
\graphicspath{{./figures/}}
\DeclareGraphicsExtensions{.pdf,.jpeg,.png}

\usepackage{url}

\usepackage[hidelinks]{hyperref}

\usepackage{balance}
\usepackage{listings}

\usepackage[absolute]{textpos}

\begin{document}

\begin{textblock}{15}(0.5,14.9)
{
\noindent\hrulefill

\noindent\fontsize{8pt}{8pt}\selectfont\copyright\ 2017 IEEE. Personal use of this material is permitted. Permission from IEEE must be obtained for all other uses, in any current or future media, including reprinting/republishing this material for advertising or promotional purposes, creating new collective works, for resale or redistribution to servers or lists, or reuse of any copyrighted component of this work in other works. \hspace{5pt} This is the accepted version of: M. Sul\'ir, J. Porub\"an. RuntimeSearch: Ctrl+F for a Running Program. Proceedings of the 32nd IEEE/ACM International Conference on Automated Software Engineering (ASE), IEEE, 2017, pp. 388--393. \url{http://doi.org/10.1109/ASE.2017.8115651}

}
\end{textblock}

%
\title{RuntimeSearch: Ctrl+F for a Running Program}

\author{
\IEEEauthorblockN{Mat\'u\v{s} Sul\'ir, Jaroslav Porub\"an}
\IEEEauthorblockA{Technical University of Ko\v{s}ice, Slovakia\\
Email: \{matus.sulir, jaroslav.poruban\}@tuke.sk}}


\maketitle

\begin{abstract}
Developers often try to find occurrences of a certain term in a software system. Traditionally, a text search is limited to static source code files. In this paper, we introduce a simple approach, RuntimeSearch, where the given term is searched in the values of all string expressions in a running program. When a match is found, the program is paused and its runtime properties can be explored with a traditional debugger. The feasibility and usefulness of RuntimeSearch is demonstrated on a medium-sized Java project.
\end{abstract}


\IEEEpeerreviewmaketitle

\section{Introduction}

Currently, the programmers use the debugger available in their IDE (integrated development environment) to execute and step a program and inspect its state. On the other hand, searching is typically performed on static source code, and it does not utilize dynamic information.

\subsection{Motivation}

Suppose a developer needs to answer a common question \cite{Sillito06questions}: Where in the source code is the label displayed in the UI (user interface) of a running program located?

Static source code search for the given term often does not produce desired results, since only a portion of the strings is located in the source code in a form of string constants. Dynamically generated and localized strings, user input, data obtained from other systems, and many other strings may not present in the static source code of the application \cite{Sulir16locating}. Instead, the developer should ask: Which expression contains the given string in its value at runtime?

A question like this could be possibly answered by writing custom scripts and queries in automated and scriptable debuggers such as Coca \cite{Ducasse99coca}, FrTime \cite{Marceau04dataflow} or \textsc{Expositor} \cite{Khoo13expositor}. However, we should take the reality of developers \cite{Reiss05paradox} into account: The cost of learning to use a tool must be lower than the expected benefit of its application. Ideally, the developer should use a tool immediately or only after minimal training. Even when using existing tools like a textual search, developers prefer simple queries compared to complicated, sophisticated ones \cite{Damevski16field}.

Considering the developer already found the source code location related to the UI element, she will probably want to examine it from the dynamic viewpoint -- inspect the concrete values of variables at a specific moment, view the stack frame, etc. Currently, searching and debugging are considered separate actions accomplished with different tools. However, according to multiple empirical studies, these two activities are often interleaved. For example, based on the results of a field study with professional developers, Damevski et al. \cite{Damevski16field} argue there should be better integration between code search, navigation and debugging. Wang et al. \cite{Wang11exploratory} studied developers locating relevant parts of source code; two of three search patterns the developers used encompassed running and debugging the program.

After finding and inspecting an initial point of investigation, developers often aim to find other occurrences relevant to this point. For instance, they try to determine where the data from a certain variable flow \cite{Ko05eliciting}. In theory, this is easy to accomplish using approaches such as dynamic program slicing \cite{Agrawal90dynamic}. In practice, the data can flow through multiple third-party systems and return back to the inspected application. For example, a user input can be saved to a database, then retrieved and displayed in another part of the application. Classical program slicing approaches could fail to find such a connection. Although cross-system slicing approaches emerge, they suffer from performance issues \cite{Binkley14orbs} or are technology-dependent \cite{Nguyen15cross}.

\subsection{Objective}

With the mentioned considerations in mind, we designed RuntimeSearch. It is a variation of a traditional textual search in source code, but in this case, we are searching in the values of expressions at runtime. A programmer enters a string which she wants to locate. It can be, for instance, a UI label visible in a running program for which she wants to find the source code part displaying it; or a string which she hypothesizes is a value of some unknown variable or expression. If the program is not already running, it is launched. During the runtime, all evaluated string expression values are being compared with the searched term. When a match is found, the program execution is paused and a traditional IDE debugger is opened. The programmer can explore runtime properties of the program like the call stack and current variable values. Then, she can continue with debugging, running, or search some string again.

In essence, RuntimeSearch is an extension of a traditional debugging process. In addition to standard debugging operations like Step In or Continue, a Find In Runtime action is available. This runtime-search action provides a user interface resembling a simple textual search dialog, which practically all developers are familiar with. Therefore, the tool should require almost no training.

A preliminary implementation is available and a case study was performed to show the feasibility and usefulness of the technique. The source code of RuntimeSearch is available online\footnote{\url{https://github.com/sulir/runtimesearch}}.

\section{Searching in the Runtime}

First, we will describe RuntimeSearch from the user's (programmer's) point of view. Next, we will look at its design and implementation.

\subsection{User's View}

The interaction and user interface of RuntimeSearch was modeled with two principles in mind:

\begin{enumerate}
\item To look similar to a traditional textual search in source code whenever it is possible. Since code searching is a familiar operation for practically every programmer, this should make learning the tool easy.
\item To integrate the approach with the debugging infrastructure already present in IDEs and used by developers.
\end{enumerate}

Each runtime searching begins by triggering the action ``Find in Runtime'' in an IDE, which shows a query prompt, where the developer enters the searched text. Subsequently, the searching process is started in one of three ways:

\begin{itemize}
\item If no program is currently being debugged, a new instance is launched.
\item If a debugged program is paused, it is resumed.
\item If a debugged program is running, it is left running.
\end{itemize}

The string is searched from the moment it was entered into a prompt until a match is found. When this happens, the program is immediately paused using a programmatically invoked breakpoint. This causes the IDE to highlight the currently executed line and show current variable values. Furthermore, all debugging features of the given IDE are available -- including the stack frames view, expression evaluation, advanced object state inspections, etc.

When the programmer considers the current search result irrelevant or wants to find the next occurrence, she triggers the action ``Find Next in Runtime''. In case she wants to change the search string, she chooses the action ``Find in Runtime''. The process continues as already described.

If the developer does not want to search a string anymore, she can continue debugging with traditional operations like Step In and Step Over or resume the program. At any moment, it is possible to search the string in the runtime if desirable.

\begin{figure}
\centering
\includegraphics[width=0.9\hsize]{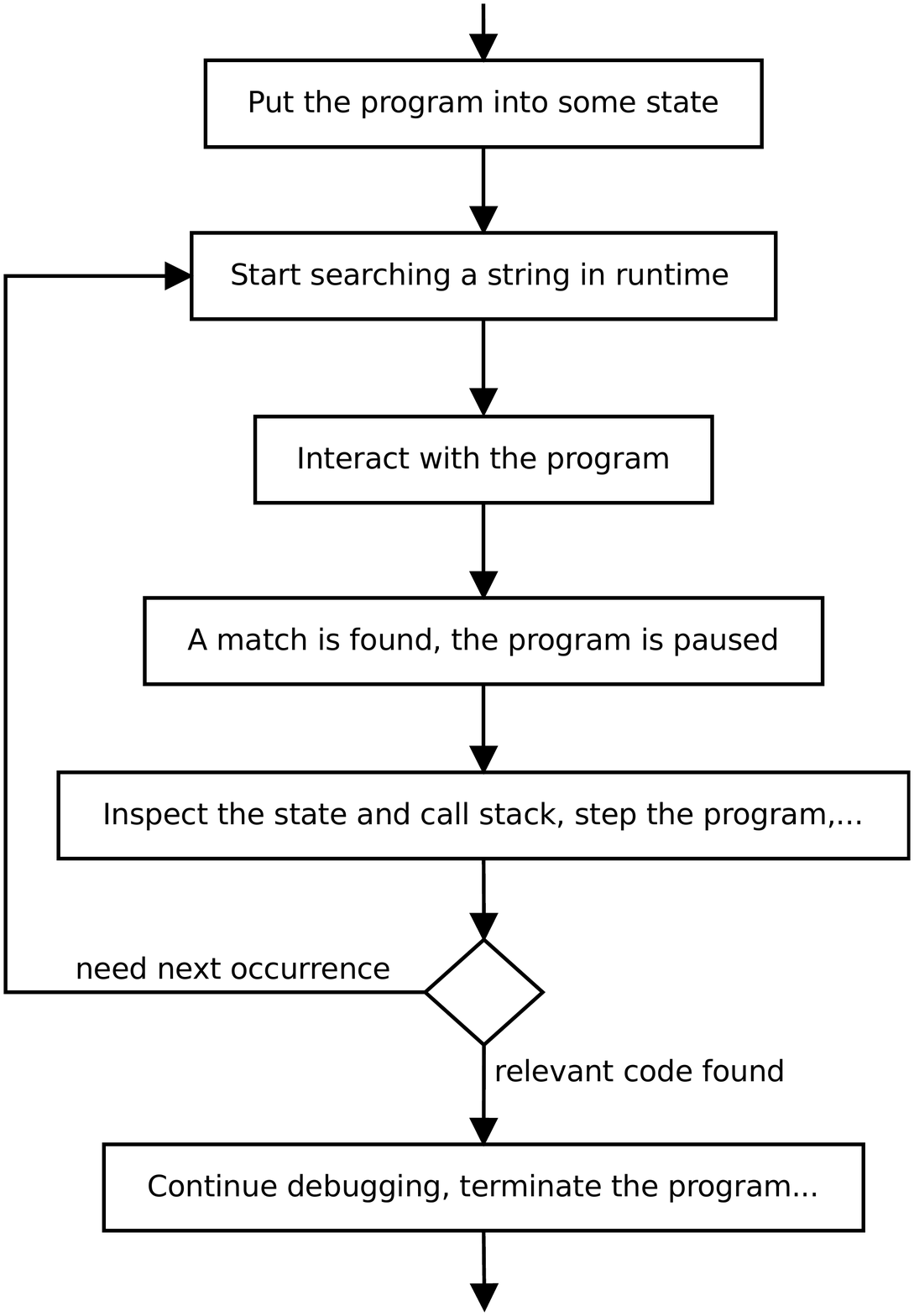}
\caption{A possible example of RuntimeSeach utilization.}
\label{f:overview}
\end{figure}

One of many possible examples of the developer's interaction with RuntimeSearch is displayed in Fig.~\ref{f:overview}. However, the process is not prescriptive and can be adapted to developer's specific needs. In section~\ref{s:study}, we will describe multiple usage scenarios.

\subsection{Principle}

We look at a running program as a series of expression evaluations. Consider the following Java source code excerpt:
{\small\begin{verbatim}
String var = "text";
var = var.toUpperCase();
\end{verbatim}}

It produces the following expression values, in the given order: ``text'' (a string constant), ``text'' (the value of the variable {\small\texttt{var}} read on the second line), and ``TEXT'' (the return value of the method {\small\texttt{toUpperCase}}).

The program is instrumented, so the result of every string expression is captured and compared to the searched text. Currently, only expressions of type {\small\texttt{String}} are captured -- it makes the most sense to compare the searched text only to strings. However, the approach is not limited to this behavior and in the future, we could convert all objects to their string representations using a {\small\texttt{toString}}-like method.

Namely, we capture the following {\small\texttt{String}} expressions: constants, local variables, member variables, constructor calls, and method calls returning values.

In our early implementation, the evaluated strings are matched against the query using a simple string containment: if the evaluated string contains the searched text, a match is found. Again, this is not an inherent limitation of the approach. It can be easily extended with standard text-search options like ``match case'', ``whole words only'', regular expression matching, or advanced features like approximate (fuzzy) string matching.

\subsection{Implementation Details}

Our RuntimeSearch implementation consists of a Java agent (a piece of instrumentation code which can be attached to any Java program) and a plugin for the IntelliJ IDEA IDE.

The agent is configurable through an argument -- a pattern specifying what packages or classes should be instrumented. This way, it is possible to specify, e.g., whether to include only application or also system classes.

Instrumentation is performed at the bytecode level. In stack-based virtual machines like the Java Virtual Machine, each expression evaluation is represented by an instruction pushing a value on the operand stack, which can be used with advantage.

The IDE plugin is very lightweight and consists mainly of a simple form and a module for communication with the agent.

\section{Case Study}
\label{s:study}

Now we will show how searching in the runtime can be useful to perform navigation and debugging tasks. The demonstration will be performed on Weka\footnote{\url{https://sourceforge.net/projects/weka/}} -- an open source Java machine learning software. It is a Swing GUI (graphical user interface) application with approximately 350,000 lines of code. A short video with parts of the case study is available online\footnote{\url{https://sulir.github.io/runtimesearch}}.

\subsection{Finding an Initial Point}

One of the questions programmers ask during program maintenance tasks is ``Where in the code is the text in this error message or UI element?'' \cite{Sillito06questions} Weka includes a package manager which displays a list of additional packages. Suppose we want to find the code which retrieves package names, such as ``AffectiveTweets''. Searching the static source code for the term ``AffectiveTweets'' does not give any results, since this string was generated at runtime.

Therefore, we run the application in a debug mode and wait until the main window appears. Then we trigger the ``Find in Runtime'' menu in the IDE, enter the query ``AffectiveTweets'' and press Find. After choosing the menu ``Package manager'' in Weka, the program automatically pauses and the IDE shows us the currently executed line, containing the searched string expression. We see the first runtime occurrence of the string ``AffectiveTweets'', i.e., its origin. Not only can we notice it was read from a file input stream (it is obvious by reading the source code), but thanks to the IDE showing a string representation of the stream object at runtime, we also see its path in the file system -- the name of the ``package list'' file from which the string ``AffectiveTweets'' was read.

\subsection{Searching for Occurrences}

After quickly finding the initial point of investigation, it is up to us how to continue. We can explore the source code, find static references, or debug the program. Another interesting possibility is to search for all following runtime occurrences of the string ``AffectiveTweets''. This can be achieved by repeatedly triggering the action ``Find Next in Runtime'' in the IDE (e.g., using a shortcut) -- each time, a new occurrence is found.

This way, we find many precise locations in source code relevant to package name retrieval and displaying. This includes reading a package list file, manipulation with an object representing the given package, reading a version list file, GUI code displaying the package name, and various helper methods. After each step, we are free to explore the current runtime properties of the program to improve our understanding, or step into a method of interest. For example, if we are interested how the version file URL is determined, we can step into the {\small\texttt{getConnection}} method while we are in the method  {\small\texttt{getRepositoryPackageVersions(String packageName)}}.

\subsection{The Fabricated Text Technique}

Instead of just ``passively'' searching for a string which the application itself displayed, we can utilize an interesting technique: Enter a made-up string into a text field and search for its runtime occurrences. This allows us to track the data flow of the string across program layers: from presentation through model to persistence.

For instance, we open the Weka Bayes Network Editor, start searching for a fabricated string like ``Node987'' in the runtime using RuntimeSearch, and create a new node named ``Node987'' in the Bayes Net editor. The debugger will immediately pause at the GUI code processing the node name. By searching for next occurrences, we are being navigated through more or less specific node-processing and data structure classes. After pressing the Save button in Weka and continuing the search, we are navigated through methods converting Bayes network nodes to XML representation and, finally, to the file-saving method.

\subsection{Non-GUI Strings}

Until now, we were searching only for strings present in the GUI in some way. However, RuntimeSearch is not limited to such texts.

We created a simple layout in the Weka KnowledgeFlow Environment, containing a DataGrid connected to a Database Server, all with default settings. Weka KnowledgeFlow contains an option to save the layout file in the KFML format (an XML dialect). We tried it, but it does not behave as expected -- after selecting this option, a file is saved as JSON (JavaScript Object Notation) instead.

\begin{figure}
\centering
\includegraphics[width=0.91\hsize]{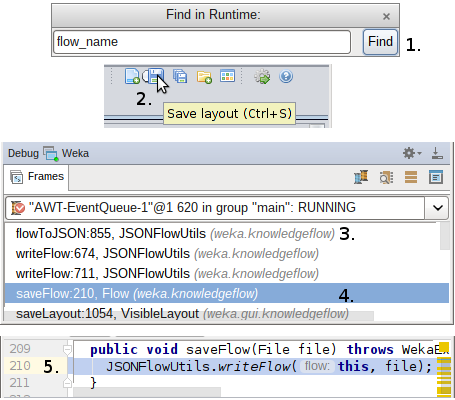}
\caption{RuntimeSearch in action: The IDE plugin (1.), an action in Weka (2.), IntelliJ IDEA debug window and source code view (3.--5.). See the main text for full description.}
\label{f:screenshot}
\end{figure}

We view the content of the file in a text editor and choose an excerpt, such as the string ``flow\_name''. We search for this excerpt in the runtime using RuntimeSearch (Fig.~\ref{f:screenshot}, part 1). After triggering the Save action in Weka (part 2 of Fig.~\ref{f:screenshot}), the program is automatically paused. We inspect the call stack: three methods at the top are located in a class starting with ``JSON'' (Fig.~\ref{f:screenshot}, label 3). This means during the execution of these methods, the wrong output format is already selected. Therefore, we click on the fourth item (label 4 in the figure). Immediately, we see the cause of the bug: The method {\small\texttt{saveFlow}} contains a hard-coded call to {\small\texttt{JSONFlowUtils.writeFlow}} (see part 5 of Fig.~\ref{f:screenshot}).

This is a demonstration how RuntimeSearch facilitates finding where in the project is the layout-saving code located, while immediately providing a context for debugging (the call stack inspection and subsequent root cause identification).

\subsection{Hypothesis Confirmation}

The strings searched so far were present either in the GUI or in external resources like files. RuntimeSearch can go even further. During debugging, developers often form hypotheses about program behavior \cite{Layman13debugging}. If the programmer has a hypothesis about a presence of a certain string in a specific member variable, she can utilize field watchpoints in common IDEs. However, if she does not know which variable is concerned, or even whether it is a variable and not only a temporary expression, runtime searching is very convenient.

If we want to open the Weka package manager from the main menu while there is no network connection available, nothing visible happens -- even no error message is displayed in the GUI, which is certainly not user-friendly. We hypothesize Weka is trying to establish an HTTP connection to load the package list, but it fails. We search for the string ``http://'' using RuntimeSearch and click the Package Manager menu item in Weka. The debugger pauses at the URL creation code. We find out the first part of our hypothesis is confirmed: An HTTP connection is created and tried to be opened. After a few ``Step Over'' actions in the IDE, we see a thrown {\small\texttt{UnknownHostException}} is caught, the stack trace is printed to the standard error stream, but no GUI message window is shown.

Similar hypotheses could be formed about SQL queries in information systems, regular expressions in parsers, etc.

\section{Performance}

As a form of a primitive benchmark, we measured the execution time of all unit tests from the package ``weka.core'' under three circumstances: without instrumentation, with instrumentation but without searching, and while searching for a short string during the whole execution. The execution was performed in debug mode, and system classes were not instrumented. A mean of three measurements was computed for each condition.

The difference between the plain and instrumented execution was negligible (14.653 vs. 14.908 s). Searching incurred reasonable overhead (38\%), the measured time was 20.223 s.

The most prominent slowdown was noticed during the ``test instantiation'' phase, not included in the above times. During the instrumented run, it lasted 8-9x more than in the plain one. This can be attributed to high overhead of the class instrumentation process itself. Note that this slowdown predominantly affects startup time, further interaction with the application is swift. Furthermore, this is only an implementation issue, and using an alternative instrumentation library should improve the performance significantly.

\section{Related Work}

Related work includes unconventional code search techniques, advanced or automated debuggers, and concept location approaches.

\subsection{Searching}

A tool by Michail \cite{Michail02browsing} builds a database containing the associations between messages shown in GUI widgets and their callbacks (along with other related functions). Then a programmer can search for a message, and associated functions are shown -- or vice versa. In contrast to RuntimeSearch, their approach is static and requires separate support for each GUI framework. Furthermore, our tool is not limited to GUI messages.

Holmes and Notkin \cite{Holmes10enhancing} describe an approach when the ``find references'' capability of an IDE is filtered using information from dynamic analysis -- only methods executed in the given scenario are returned. Compared to RuntimeSearch, they do not capture nor search in the values of expressions.

\textsc{Spotter} \cite{Chis16moldable} is a framework for the creation of custom search processors in the Pharo environment. Although such search processors have access also to runtime data, and in theory, a tool similar to RuntimeSearch could be built, no such search processor was described in the article.

DynamiDoc \cite{Sulir17generating} writes Javadoc documentation above methods, directly into source code files. It contains concrete examples of parameters, return values and state changes during execution. It would be possible to use standard text search to find the given value in the generated documentation.

\subsection{Debugging}

Automated debuggers like Coca \cite{Ducasse99coca} or a scriptable debugger by Marceau et al. \cite{Marceau04dataflow} perceive an executing program as a set of events. They allow developers to write predicates describing when the program should pause and scripts automatically performing given actions. \textsc{Expositor} \cite{Khoo13expositor} adds time-traveling capabilities to such scripts. Although these and similar systems are more feature-rich than RuntimeSearch, we believe the simplicity of our approach will allow developers to learn it quickly, which should accelerate its possible industrial adoption. Instead of writing scripts and queries, RuntimeSearch offers a familiar user interface of a text search dialog and applies it to runtime.

Whyline \cite{Ko08debugging} allows developers to ask Why and Why Not questions about the observed program behavior. For example, we can click on a drawn square and ask why it has a specific color. The relevant piece of code is shown, along with a sequence of steps how the color was computed. Compared to RuntimeSearch, Whyline operates not only with strings but also with graphical elements in a program. However, it has multiple limitations. First, it is offline, i.e., trace-based, which seriously limits its practical adoption because of a massive amount of data collected \cite{Ko08debugging}. Second, it substitutes, not complements the built-in IDE debugger which many developers are already used to. Third, it uses identifier names to suggest possible questions, so it relies heavily on good naming.

With object-centric debugging \cite{Ressia12object}, we can place breakpoints on specific object instances at runtime. The breakpoints are then triggered when this instance is manipulated in a given way. For example, we could track a specific string instance in a running program. However, to use object-centric debugging, the program must be already paused and a specific object instance must be manually selected. RuntimeSearch automatically finds such an instance.

\subsection{Concept Location}

In general, concept location (or feature location) is the process of finding where the given concept or feature is implemented in the source code \cite{Rajlich02role}. Feature location approaches take a high-level description of a feature as an input, and they produce a list of code elements contributing to this functionality. In this sense, RuntimeSearch is not a feature location approach. On the other hand, one of the possible uses of our tool is to locate UI terms in the values of expressions in the source code. Therefore, we can consider RuntimeSearch an auxiliary tool useful in the concept location process.

Chen and Rajlich \cite{Chen00case} present feature location as a guided search across the program's abstract system dependence graph. Starting from a selected program element, the developer regulates the feature location process by choosing relevant nodes, while the computer controls the traversal. Similar to RuntimeSearch, the technique is interactive and alternates the control between a user and a computer. However, it is purely static and it does not utilize information from dynamic analysis.

Bohnet and D\"ollner \cite{Bohnet08analyzing} describe a feature location approach where execution traces are collected and visualized in a form of a call graph. Then, the developer inspects the call graph and selects specific points of interest. In a subsequent run, values of parameters or variables are collected at the specific points, and the graph and source code is annotated with them. Compared to RuntimeSearch, their approach requires two separate instrumented executions. In their method, only a small, manually selected portion of variable values is collected. Finally, their tool is not integrated with an IDE and its ready-to-use debugging capabilities.

Some feature location approaches (e.g., SITIR \cite{Liu07feature}) combine dynamic analysis with information retrieval applied on the source code -- this relies on the programmers using proper identifiers. In contrast to them, we utilize the values of string variables and expressions at runtime.

A feature location approach by Anwikar et al. \cite{Anwikar12domain} utilizes concrete variable values in the analysis, but only in a very limited way. First, the approach uses partial evaluation, and no real program execution is performed. Second, the approach is limited to ``function variables'' in legacy software -- variables which determine the type of an action to be performed, containing one of a predefined set of values.

FLAT$^3$ \cite{Savage10flat3} and I3 \cite{Beck15rethinking} are advanced user interfaces for feature location. Instead of providing a custom GUI, RuntimeSearch tries to utilize existing IDE capabilities as much as possible.

One of the open problems of dynamic feature location is the selection of appropriate program inputs. Hayashi et al. \cite{Hayashi16guiding} propose a technique to guide the identification of unexplored scenarios. Since our technique does not aim for a complete feature-code mapping, only for temporary queries, scenario selection is much less of a concern. A developer executes a program according to a scenario (s)he is interested in, and the results are specific to this execution.

\section{Conclusion and Future Work}

In this paper, we presented RuntimeSearch -- a simple search engine which, instead of searching in the static source code, searches in the values of all string expressions of a running program. It applies a well-known ``find text'' metaphor in the runtime. Our tool integrates into the debugging infrastructure of an IDE and extends its capabilities.

In a case study on a 350 kLOC open source system, we have shown how it can be used to locate an initial investigation point in the program by finding the location displayed in the GUI. Next, we continued to find other occurrences of the same string to find more (possibly) related pieces of code. Our tool is helpful also during debugging, when the programmer hypothesizes about the presence of a certain string in an unknown variable.

The currently implemented version of RuntimeSearch is very limited. The first future research area is an improvement of the matching options, e.g. regular expression or fuzzy string matching. Ideally, a programmer would be able to enter a high-level description of a feature and the program would pause when the feature is executed, which could be regarded as ``feature breakpoints''.

Sometimes we encountered a situation when multiple ``Find Next in Runtime'' operations in a row pointed to the same statement. This happened mainly in loops which contained the searched string expressions in their body. Adding an option to automatically skip such occurrences is an interesting future work idea. Similarly, we could limit the search granularity by pausing only at one occurrence during a method call.

Storing previous occurrences and thus enabling the ``Find Previous'' operation could be useful too. Of course, the previous occurrences would be only static source code locations, since storing runtime state would effectively mean building a time-traveling debugger.

Searching only for {\small\texttt{String}} objects is limiting. Extension to numeric values is a viable option. In many languages, including Java, it is possible to convert an object to its string representation using a method like {\small\texttt{toString()}}. This could be used to perform a text search in various non-string objects.

In a small benchmark, the overhead of running instrumented code without an active search was negligible after the start-up (class-loading) period. Searching for a string incurred and overhead of about 38\%. Optimization and a thorough performance evaluation are planned.
 
Finally, we plan to fully validate the approach using a controlled experiment with human participants performing software comprehension and maintenance tasks.

\section*{Acknowledgment}

This work was supported by project KEGA 047TUKE-4/2016 Integrating software processes into the teaching of programming.

\bibliographystyle{./IEEEtran}
\bibliography{ase}

\begin{thebibliography}{10}
\providecommand{\url}[1]{#1}
\csname url@samestyle\endcsname
\providecommand{\newblock}{\relax}
\providecommand{\bibinfo}[2]{#2}
\providecommand{\BIBentrySTDinterwordspacing}{\spaceskip=0pt\relax}
\providecommand{\BIBentryALTinterwordstretchfactor}{4}
\providecommand{\BIBentryALTinterwordspacing}{\spaceskip=\fontdimen2\font plus
\BIBentryALTinterwordstretchfactor\fontdimen3\font minus
  \fontdimen4\font\relax}
\providecommand{\BIBforeignlanguage}[2]{{%
\expandafter\ifx\csname l@#1\endcsname\relax
\typeout{** WARNING: IEEEtran.bst: No hyphenation pattern has been}%
\typeout{** loaded for the language `#1'. Using the pattern for}%
\typeout{** the default language instead.}%
\else
\language=\csname l@#1\endcsname
\fi
#2}}
\providecommand{\BIBdecl}{\relax}
\BIBdecl

\bibitem{Sillito06questions}
J.~Sillito, G.~C. Murphy, and K.~De~Volder, ``Questions programmers ask during
  software evolution tasks,'' in \emph{Proceedings of the 14th ACM SIGSOFT
  International Symposium on Foundations of Software Engineering}, ser. SIGSOFT
  '06/FSE-14.\hskip 1em plus 0.5em minus 0.4em\relax New York, NY, USA: ACM,
  2006, pp. 23--34.

\bibitem{Sulir16locating}
M.~Sul\'ir and J.~Porub\"an, ``Locating user interface concepts in source
  code,'' in \emph{5th Symposium on Languages, Applications and Technologies
  (SLATE'16)}, ser. OpenAccess Series in Informatics (OASIcs), vol.~51.\hskip
  1em plus 0.5em minus 0.4em\relax Dagstuhl, Germany: Schloss
  Dagstuhl--Leibniz-Zentrum fuer Informatik, 2016, pp. 6:1--6:9.

\bibitem{Ducasse99coca}
M.~Ducasse, ``Coca: an automated debugger for {C},'' in \emph{Proceedings of
  the 1999 International Conference on Software Engineering}, May 1999, pp.
  504--513.

\bibitem{Marceau04dataflow}
G.~Marceau, G.~H. Cooper, S.~Krishnamurthi, and S.~P. Reiss, ``A dataflow
  language for scriptable debugging,'' in \emph{Proceedings. 19th International
  Conference on Automated Software Engineering, 2004.}, Sept 2004, pp.
  218--227.

\bibitem{Khoo13expositor}
Y.~P. Khoo, J.~S. Foster, and M.~Hicks, ``Expositor: Scriptable time-travel
  debugging with first-class traces,'' in \emph{Proceedings of the 2013
  International Conference on Software Engineering}, ser. ICSE '13.\hskip 1em
  plus 0.5em minus 0.4em\relax Piscataway, NJ, USA: IEEE Press, 2013, pp.
  352--361.

\bibitem{Reiss05paradox}
S.~P. Reiss, ``The paradox of software visualization,'' in \emph{Proceedings of
  the 3rd IEEE International Workshop on Visualizing Software for Understanding
  and Analysis}, ser. VISSOFT '05.\hskip 1em plus 0.5em minus 0.4em\relax
  Washington, DC, USA: IEEE Computer Society, 2005, pp. 59--63.

\bibitem{Damevski16field}
K.~Damevski, D.~Shepherd, and L.~Pollock, ``A field study of how developers
  locate features in source code,'' \emph{Empirical Software Engineering},
  vol.~21, no.~2, pp. 724--747, 2016.

\bibitem{Wang11exploratory}
J.~Wang, X.~Peng, Z.~Xing, and W.~Zhao, ``An exploratory study of feature
  location process: Distinct phases, recurring patterns, and elementary
  actions,'' in \emph{Proceedings of the 2011 27th IEEE International
  Conference on Software Maintenance}, ser. ICSM '11.\hskip 1em plus 0.5em
  minus 0.4em\relax Washington, DC, USA: IEEE Computer Society, 2011, pp.
  213--222.

\bibitem{Ko05eliciting}
A.~J. Ko, H.~H. Aung, and B.~A. Myers, ``Eliciting design requirements for
  maintenance-oriented {IDE}s: A detailed study of corrective and perfective
  maintenance tasks,'' in \emph{Proceedings of the 27th International
  Conference on Software Engineering}, ser. ICSE '05.\hskip 1em plus 0.5em
  minus 0.4em\relax New York, NY, USA: ACM, 2005, pp. 126--135.

\bibitem{Agrawal90dynamic}
H.~Agrawal and J.~R. Horgan, ``Dynamic program slicing,'' \emph{SIGPLAN Not.},
  vol.~25, no.~6, pp. 246--256, Jun. 1990.

\bibitem{Binkley14orbs}
D.~Binkley, N.~Gold, M.~Harman, S.~Islam, J.~Krinke, and S.~Yoo, ``O{RBS}:
  Language-independent program slicing,'' in \emph{Proceedings of the 22Nd ACM
  SIGSOFT International Symposium on Foundations of Software Engineering}, ser.
  FSE 2014.\hskip 1em plus 0.5em minus 0.4em\relax New York, NY, USA: ACM,
  2014, pp. 109--120.

\bibitem{Nguyen15cross}
H.~V. Nguyen, C.~K\"{a}stner, and T.~N. Nguyen, ``Cross-language program
  slicing for dynamic web applications,'' in \emph{Proceedings of the 2015 10th
  Joint Meeting on Foundations of Software Engineering}, ser. ESEC/FSE
  2015.\hskip 1em plus 0.5em minus 0.4em\relax New York, NY, USA: ACM, 2015,
  pp. 369--380.

\bibitem{Layman13debugging}
L.~Layman, M.~Diep, M.~Nagappan, J.~Singer, R.~DeLine, and G.~Venolia,
  ``Debugging revisited: Toward understanding the debugging needs of
  contemporary software developers,'' in \emph{2013 ACM / IEEE International
  Symposium on Empirical Software Engineering and Measurement}.\hskip 1em plus
  0.5em minus 0.4em\relax Los Alamitos, CA, USA: IEEE Computer Society, 2013,
  pp. 383--392.

\bibitem{Michail02browsing}
A.~Michail, ``Browsing and searching source code of applications written using
  a {GUI} framework,'' in \emph{Proceedings of the 24th International
  Conference on Software Engineering}, ser. ICSE '02.\hskip 1em plus 0.5em
  minus 0.4em\relax New York, NY, USA: ACM, 2002, pp. 327--337.

\bibitem{Holmes10enhancing}
R.~Holmes and D.~Notkin, ``Enhancing static source code search with dynamic
  data,'' in \emph{Proceedings of 2010 ICSE Workshop on Search-driven
  Development: Users, Infrastructure, Tools and Evaluation}, ser. SUITE
  '10.\hskip 1em plus 0.5em minus 0.4em\relax New York, NY, USA: ACM, 2010, pp.
  13--16.

\bibitem{Chis16moldable}
A.~Chi\c{s}, T.~G\^{\i}rba, J.~Kubelka, O.~Nierstrasz, S.~Reichhart, and
  A.~Syrel, ``Moldable, context-aware searching with {S}potter,'' in
  \emph{Proceedings of the 2016 ACM International Symposium on New Ideas, New
  Paradigms, and Reflections on Programming and Software}, ser. Onward!
  2016.\hskip 1em plus 0.5em minus 0.4em\relax New York, NY, USA: ACM, 2016,
  pp. 128--144.

\bibitem{Sulir17generating}
M.~Sul\'ir and J.~Porub\"an, ``Generating method documentation using concrete
  values from executions,'' in \emph{6th Symposium on Languages, Applications
  and Technologies (SLATE 2017)}, ser. OpenAccess Series in Informatics
  (OASIcs), vol.~56.\hskip 1em plus 0.5em minus 0.4em\relax Dagstuhl, Germany:
  Schloss Dagstuhl--Leibniz-Zentrum fuer Informatik, 2017, pp. 3:1--3:13.

\bibitem{Ko08debugging}
A.~J. Ko and B.~A. Myers, ``Debugging reinvented: Asking and answering why and
  why not questions about program behavior,'' in \emph{Proceedings of the 30th
  International Conference on Software Engineering}, ser. ICSE '08.\hskip 1em
  plus 0.5em minus 0.4em\relax New York, NY, USA: ACM, 2008, pp. 301--310.

\bibitem{Ressia12object}
J.~Ressia, A.~Bergel, and O.~Nierstrasz, ``Object-centric debugging,'' in
  \emph{Proceedings of the 34th International Conference on Software
  Engineering}, ser. ICSE '12.\hskip 1em plus 0.5em minus 0.4em\relax
  Piscataway, NJ, USA: IEEE Press, 2012, pp. 485--495.

\bibitem{Rajlich02role}
V.~Rajlich and N.~Wilde, ``The role of concepts in program comprehension,'' in
  \emph{Program Comprehension, 2002. Proceedings. 10th International Workshop
  on}, 2002, pp. 271--278.

\bibitem{Chen00case}
K.~Chen and V.~Rajlich, ``Case study of feature location using dependence
  graph,'' in \emph{Proceedings of the 8th International Workshop on Program
  Comprehension}, ser. IWPC '00.\hskip 1em plus 0.5em minus 0.4em\relax
  Washington, DC, USA: IEEE Computer Society, 2000, pp. 241--247.

\bibitem{Bohnet08analyzing}
J.~Bohnet and J.~D\"ollner, ``Analyzing dynamic call graphs enhanced with
  program state information for feature location and understanding,'' in
  \emph{Companion of the 30th International Conference on Software
  Engineering}, ser. ICSE Companion '08.\hskip 1em plus 0.5em minus 0.4em\relax
  New York, NY, USA: ACM, 2008, pp. 915--916.

\bibitem{Liu07feature}
D.~Liu, A.~Marcus, D.~Poshyvanyk, and V.~Rajlich, ``Feature location via
  information retrieval based filtering of a single scenario execution trace,''
  in \emph{Proceedings of the Twenty-second IEEE/ACM International Conference
  on Automated Software Engineering}, ser. ASE '07.\hskip 1em plus 0.5em minus
  0.4em\relax New York, NY, USA: ACM, 2007, pp. 234--243.

\bibitem{Anwikar12domain}
V.~Anwikar, R.~Naik, A.~Contractor, and H.~Makkapati, ``Domain-driven technique
  for functionality identification in source code,'' \emph{ACM SIGSOFT Software
  Engineering Notes}, vol.~37, no.~3, pp. 1--8, May 2012.

\bibitem{Savage10flat3}
T.~Savage, M.~Revelle, and D.~Poshyvanyk, ``F{LAT}$^3$: Feature location and
  textual tracing tool,'' in \emph{Proceedings of the 32nd ACM/IEEE
  International Conference on Software Engineering - Volume 2}, ser. ICSE
  '10.\hskip 1em plus 0.5em minus 0.4em\relax New York, NY, USA: ACM, 2010, pp.
  255--258.

\bibitem{Beck15rethinking}
F.~Beck, B.~Dit, J.~Velasco-Madden, D.~Weiskopf, and D.~Poshyvanyk,
  ``Rethinking user interfaces for feature location,'' in \emph{Proceedings of
  the 2015 IEEE 23rd International Conference on Program Comprehension}, ser.
  ICPC '15.\hskip 1em plus 0.5em minus 0.4em\relax Piscataway, NJ, USA: IEEE
  Press, 2015, pp. 151--162.

\bibitem{Hayashi16guiding}
S.~Hayashi, H.~Kazato, T.~Kobayashi, T.~Oshima, K.~Natsukawa, T.~Hoshino, and
  M.~Saeki, ``Guiding identification of missing scenarios for dynamic feature
  location,'' in \emph{2016 23rd Asia-Pacific Software Engineering Conference
  (APSEC)}, Dec. 2016, pp. 393--396.

\end{thebibliography}
\balance

\end{document}